\documentclass[12pt,a4paper]{article}
\usepackage{graphicx,amsmath}
\usepackage{url}
\newcommand{\be}{\begin{equation}}
\newcommand{\ee}{\end{equation}}
\newcommand{\ben}{\begin{equation*}}
\newcommand{\een}{\end{equation*}}
\newcommand{\bea}{\begin{eqnarray}}
\newcommand{\eea}{\end{eqnarray}}
\newcommand{\ar}{\begin{array}}
	\newcommand{\arn}{\end{array}}

\def\pnot{\mbox{${\not{\hbox{\kern-3.0pt$p$}}}$}}
\def\qnot{\mbox{${\not{\hbox{\kern-2.0pt$q$}}}$}}
\def\enot{\mbox{${\not{\hbox{\kern-2.0pt$e$}}}$}}
\def\knot{\mbox{${\not{\hbox{\kern-2.0pt$k$}}}$}}

\def\fun#1#2{\lower3.6pt\vbox{\baselineskip0pt\lineskip.9pt\ialign
		{$\mathsurround=0pt#1\hfil##\hfil$\crcr#2\crcr\sim\crcr}}}

\makeatletter 
\def\appendix{\par\clearpage 
	\setcounter{section}{0} 
	\setcounter{subsection}{0} 
	\@addtoreset{equation}{section} 
	\def\@sectname{Appendix~} 
	\def\theequation{\thesection.\arabic{equation}} 
	\def\thesection{\Alph{section}}} 
\makeatother 

\begin{document}

\bigskip
\bigskip
\begin{center}
{\bf\Large The Low soft-photon theorem again }\\
 
 \vspace{1cm}
   
 V.~S.~Fadin$^{a}$ and  V.~A.~Khoze$^{b}$ \\
 
 \vspace{0.5cm}

 $^a$ Budker Institute of Nuclear Physics, Novosibirsk 630090, Russia\\
 $^b$ Institute for Particle Physics Phenomenology, Durham University, DH1 3LE, UK\\

 \vspace{1cm}

 \abstract{\noindent It is shown that contrary to claims of Ref.\cite {Lebiedowicz:2021byo} the formulated 
in the proper physical variables Low theorem \cite{Low:1958sn} for soft photon emission
does not require any modification.
We also reject the criticism in Ref.\cite {Lebiedowicz:2021byo} of the papers  \cite{Burnett:1967km, Lipatov:1988ii}.
At the same time, we identify some inaccuracies in  Ref.\cite{Burnett:1967km} in the presentation of the soft-photon theorem 
for the case of spin-one-half particles.
We also point out shortcomings in consideration of the Low theorem in the classic textbooks \cite{Berestetskii:1982qgu, BLP:1971}.
 }\\

\vfill

 E-mail:  \url{v.s.fadin@inp.nsk.su},
 \url{v.a.khoze@durham.ac.uk}

 \end{center}

 \newpage

\section{Introduction}

Nearly 65 years ago in his celebrated paper \cite{Low:1958sn}
Francis Low  proved that the first two terms in the series  expansion 
of the differential radiative cross-section 
in  powers of photon energy $k$ 
can be expressed via the corresponding non-radiative amplitude.
  
This result was extended to the cross-sections of unpolarized multiparticle processes
involving charged 
spin 1/2 particles by Burnett and Kroll in \cite{Burnett:1967km}
and generalized subsequently by Bell and Van Royen \cite{Bell:1969yw}  to particles of arbitrary spin.
The studies of soft photon radiation have a very long history and 
has continued to attract attention until now, see  \cite{Lebiedowicz:2021byo},
\cite{Lebiedowicz:2023mlz}~-~\cite{Balsach:2024rkn} and references therein.
Recently the authors of  \cite{Lebiedowicz:2021byo}(see also \cite{Lebiedowicz:2023mlz,Lebiedowicz:2023ell})
have questioned the Low theorem, in what concerns
the validity of the second term in the expansion over photon momentum.
They also disputed some results of the basic works  \cite{Burnett:1967km, Lipatov:1988ii}.

This paper aims to demonstrate that the critical comments of \cite{Lebiedowicz:2021byo}
concerning the consistency of the Low theorem and some results of  \cite{Burnett:1967km, Lipatov:1988ii}
are unjustified \footnote{Errors in  interpretation of the  Low
theorem in  \cite{Lebiedowicz:2021byo}
were also pointed out in \cite{Balsach:2023ema,Balsach:2024rkn}. }.
While addressing this issue we found some inaccuracies in \cite{Burnett:1967km} and in
the presentation of the Low theorem in the popular textbooks  \cite{Berestetskii:1982qgu, BLP:1971}.

In the next section, we consider the soft-photon theorem for the case of two spin-zero particles 
exemplified similarly to \cite{Lebiedowicz:2021byo} by the radiation in the $\pi^- \pi^0$ scattering.
In section 3 we address the Low theorem for the case of the process with spin-one-half emitter, and in 
section 4 we consider the  Burnett-Kroll extension of the soft photon theorem.
Some inconsistencies in the presentation of the soft theorem in \cite{Berestetskii:1982qgu, BLP:1971} are discussed in section 5 and
we conclude in section 6.

\section{The  Low theorem for scalar particles}
Let us use   for particle momenta the same notation as in Ref. \cite{Lebiedowicz:2021byo}  and consider the  
 processes            
\be
\pi^-(p_a)+\pi^0(p_b)\rightarrow \pi^-(p_1)+\pi^0(p_2) \label{elastic}
\ee
with 
\be
p_a+p_b = p_1 +p_2, \;\; p_a^2=p_1^2 = m_a^2, \;\;  p_b^2=p_2^2 = m_b^2, \label{elastic momenta}
\ee
and 
\be
\pi^-(p_a)+\pi^0(p_b)\rightarrow \pi^-(p'_1)+\pi^0(p'_2) +\gamma (k, e) \label{inelastic}
\ee
with 
\be
p_a+p_b = p'_1 +p'_2 +k, \;\; 
p_a^2= p_1^{'2} = m_a^2, \;\; p_b^2= p_2^{'2} = m_b^2, \;\; k^2 =0~.\label{inelastic momenta}
\ee 
It is convenient to introduce notation
\be
j_\mu =\frac{p'_{1\mu}}{(p'_1k)}-\frac{p_{a\mu}}{(p_a k)}~, \;\;{\cal P}_{f}^{\mu} =t_f^{\mu\rho}p'_{2\rho}~, \;\;{\cal P}_{i}^{\mu} =t_i^{\mu\rho}p_{b\rho}~, \label{Pif}
\ee
where
\be
t_i^{\mu\rho} = \frac{p_{a}^\mu}{(p_a k)} k^\rho -g^{\mu\rho}~, \;\;
t_f^{\mu\rho} = \frac{p'^{\mu}_{1}}{(p'_1 k)}k^\rho -g^{\mu\rho}~. \label{tif}
\ee

Then the original  Low theorem \cite{Low:1958sn} reads (see Eq. (2.16) in  \cite{Low:1958sn})
\be
M_{\mu}=e\Bigg[j_\mu  +\Bigg({\cal P}_{i\mu}+{\cal P}_{f\mu}\Bigg)  \frac{\partial}{\partial \nu}\Bigg] T (\nu, \Delta)~, \label{low}
\ee
where 
\be
\nu = (p_ap_b)+(p_1'p_2')~,\;\; \Delta = (p_b-p_2')^2,  \label{nu, Delta}
\ee
and $T (\nu, \Delta) = T(p_1^{'2}, p_a^2, ;\nu, \Delta)|_{p_1^{'2}=p_a^2 = m_a^2}$,  where $T(p_1^{'2}, p_a^2, ; \nu, \Delta)$ is the off-mass-shell amplitude (amputated Green function) of  elastic  process.  
 It is clearly stated in \cite{Low:1958sn} (see Eq.(2.9) and a phrase below it) that this amplitude " conserves momentum and energy
 but not mass."  Let us emphasize that the amplitude of the process (\ref{elastic}) is  
a function of two invariant variables, $T (\nu_i, t_b)$, with $\nu_i =2(p_ap_b)$  and $t_b=(p_b-p_2)^2$. \\
 The important point here is that the Low theorem is formulated in terms of the physical momenta of the radiative process (\ref{inelastic}). 
The authors of \cite{Lebiedowicz:2021byo} claim that such consideration is incorrect because these momenta 
cannot be the momenta of the non-radiative (elastic) process. 
Below Eq.(3.40) in \cite{Lebiedowicz:2021byo} it is written  "The term of order $\omega^0$ 
\footnote{Note that in the notation of Ref. \cite{Lebiedowicz:2021byo} $\omega$ is the photon energy.}
 in the expansion
 of the amplitude given in \cite{Low:1958sn} corresponds to the \underline {fictitious} process (3.31) which does not respect
energy-momentum conservation."  
  Here (3.31) means
\be
\pi^-(p_a)+\pi^0(p_b)\rightarrow \pi^-(p_1)+\pi^0(p_2) +\gamma (k, e),  \label{inelastic 1}
\ee
i.e. the inelastic process (\ref{inelastic}), but with momenta of the final pions corresponding to the non-radiative case (\ref{inelastic}). 
However, such an equation
has never been presented in \cite{Low:1958sn}. Actually,  everywhere in \cite{Low:1958sn} the physical momenta of the process (\ref{inelastic}) satisfying the energy-momentum conservation
 (\ref{inelastic momenta}) are used. In contrast, the momenta used in  \cite{Lebiedowicz:2021byo}
correspond to the
 elastic process.
 It is true, that the momenta $p_a, p_b, p'_1, p'_2$ 
used in \cite{Low:1958sn} cannot be  momenta of  the elastic process, because they do not satisfy  energy-momentum conservation
 for this process, $p_a+p_b\neq p'_{1}+ p'_2$. But the arguments $ \nu_i$ and $t_b$ of the elastic scattering amplitude $T ( \nu_i, t_b)$ are independent variables, so, certainly,
 they  can be taken equal to
$\nu= (p_ap_b)+(p_1'p_2')$ and $\Delta= t_b =(p_b-p'_2)^2$. 
Instead of using these variables the authors of  \cite{Lebiedowicz:2021byo}
introduced
\underline {artificial} momenta $p_1$ and $p_2$ satisfying the elastic conservation law $p_a+p_b =p_1+ p_2$  and  reformulate the theorem in terms of these momenta.  The momenta $p_1$ and $p_2$ in \cite{Lebiedowicz:2021byo} are written as 
\be
p_1= p_1'+l_1~, \;\; p_2= p_2'+l_2~, \label{pi}
\ee
where the momenta $l_1$ and $l_2$ are small (${\cal O}$ ($k$)) and 
\be
l_1+l_2=k~.\label{l_i}
\ee
Since both $p_1$ and $p_1'$, as well as $p_2$ and $p_2'$ are on-shell, we get
\be
2(p'_1l_1) =-l_1^2, \;\; 2(p'_2l_2) =-l_2^2~.\label{transverse}
\ee
In Ref.\cite{Lebiedowicz:2021byo} some particular representation for the momenta $l_i$ is used, which is not essential for
the discussion below. The amplitude of the radiative process is presented there (see Eqs. (A1) or (3.27) and (3.28) 
in \cite{Lebiedowicz:2021byo})
 at $k^2 =0, (\epsilon^*k) =0 $, where $\epsilon^*$ is the photon polarization vector,  as
\[
{\cal M}_\lambda = e {\cal M}^{(0)}(s_L, t, m^2_\pi, m^2_\pi, m^2_\pi,m^2_\pi)\Bigg[\frac{p_{a\lambda}}{(p_a k)} -\frac{p_{1\lambda}}{(p_1 k)} -\frac{p_{1\lambda}(l_1k) -l_{1\lambda}(p_1k)}{(p_1k)^2}\Bigg] 
\]
\[
+ 2e\frac{\partial}{\partial s_L }{\cal M}^{(0)}(s_L, t, m^2_\pi, m^2_\pi, m^2_\pi,m^2_\pi)\Big[p_{b\lambda}-p_{a\lambda}\frac{(p_b k)}{(p_a k)} \Bigg] 
\]
\be
-2e\frac{\partial}{\partial t }{\cal M}^{(0)}(s_L, t, m^2_\pi, m^2_\pi, m^2_\pi,m^2_\pi)\Bigg[\frac{p_{a\lambda}}{(p_a k)} -\frac{p_{1\lambda}}{(p_1 k)}\Bigg]\Bigg[\left((p_a-p_1)k\right)-(p_a l_1)\Bigg]~.\label{LNS}
\ee
Here $s_L=(p_ap_b)+ (p_1p_2), \;\; t=(p_a-p_1)^2= (p_b-p_2)^2, \;\; {\cal M}^{(0)}(s_L, t, m^2_\pi, m^2_\pi, m^2_\pi,m^2_\pi) $ is the elastic scattering amplitude. 

Note that $s_L\neq \nu $ and $t\neq \Delta$. With the accuracy to order $k$  accounting for Eq. (\ref{transverse}), we have 
\[
s_L = \nu +k(p'_2+p'_1)= \nu +k(p_2+p_1)~, \;\; t=\Delta -2(l_2p_b) ~.
\]

Therefore, taking the first term in (\ref{LNS})  the expansion in $k$
\[
{\cal M}^{(0)}(s_L, t, m^2_\pi, m^2_\pi, m^2_\pi,m^2_\pi) ={\cal M}^{(0)}(\nu, \Delta, m^2_\pi, m^2_\pi, m^2_\pi,m^2_\pi) 
\]
\[
+\left(k(p_a+p_b)\right))\frac{\partial}{\partial \nu }{\cal M}^{(0)}(\nu, \Delta, m^2_\pi, m^2_\pi, m^2_\pi,m^2_\pi) 
\]
\be
-2(l_2p_b)\frac{\partial}{\partial \Delta }{\cal M}^{(0)}(\nu, \Delta, m^2_\pi, m^2_\pi, m^2_\pi,m^2_\pi) 
\ee
and setting in the second and the third terms $s_L=\nu, \;\; t=\Delta$ (that is allowed
because these terms are ${\cal O}$ ($k$)),
 we arrive at the original formulation of the Low theorem (apart from the difference in common sign of the amplitudes in \cite{Lebiedowicz:2021byo} and \cite{Low:1958sn}). 
Thus, the formulation of the Low theorem proposed in \cite{Lebiedowicz:2021byo} fully agrees with the original one \cite{Low:1958sn}. However, it uses artificially constructed variables that differ from the physical variables of the radiative process used in Ref.  \cite{Low:1958sn}).
It is worth emphasizing here that such an agreement holds just for the first two terms in the expansion in the photon energy. Only these two terms obey the theorems for soft-photon radiation based on the gauge invariance.
Let us note here that the choice in \cite{Low:1958sn}
 of $\nu$ and $\Delta$ as two independent variables of the non-radiative scattering amplitude is 
a convenient,  but obviously
not the only option (see, e.g.  footnote 3 in \cite{Burnett:1967km})
Therefore, different formulations of the theorem are possible. For example, taking as the two independent variables $\nu_i = 2(p_ap_b)$ and $\Delta$ and using the relation
\be
\nu = \nu_i -(k(p_a+p_b))~, \label{nu and nu i}  
\ee
we obtain with  the  ${\cal O}$ ($k$)  accuracy
\be
T(\nu, \Delta) = -(k(p_a+p_b))\frac{\partial}{\partial {\nu_i} }T(\nu_i, \Delta)~. \label{T nu i} 
\ee
Inserting (\ref{T nu i}) in (\ref{low})  with an account for momentum conservation  and neglecting in
the  $M_{\mu}$ terms proportional to $k_\mu$ 
we get  
\be
M_{\mu}=e\Bigg[j_\mu +2P_{i\mu} \frac{\partial}{\partial{\nu_i}}\Bigg] T (\nu_i, \Delta)~. \label{low i}
\ee
Similarly, we can obtain 
\be
M_{\mu}=e\Bigg[j_\mu +2P_{f\mu} \frac{\partial}{\partial{\nu_f}}\Bigg] T (\nu_f, \Delta)~,   \label{low f}
\ee
where $\nu_f = 2(p'_1p'_2)$. 
In the expressions (\ref{low}), (\ref{low i}), (\ref{low f}) the first arguments of the amplitude of the non-radiative process differ by  the ${\cal O}$ ($k$)
terms.
It is always possible to set the second argument also being different by the terms of the same order. For example, we can write the 
singular term as $ej_\mu T(\nu_i, \Delta_p)$, where $\Delta_p = (p_a-p'_1)^2 = \Delta +2k(p_a-p'_1)$. In this case, with the required accuracy 
\be
T(\nu_i, \Delta) =T(\nu_i, \Delta_p) -2(k(p_a-p'_1))\frac{\partial}{\partial{\Delta_p}}T(\nu_i, \Delta_p)  
\ee
and we obtain from (\ref{low i})
\be
M_{\mu}=e\Bigg[j_\mu  +2{\cal P}_{i\mu} \frac{\partial}{\partial \nu} -2\big({\cal R}_{i\mu}+{\cal R}_{f\mu}\big)  \frac{\partial}{\partial {\Delta_p}}\Bigg] T (\nu_i, \Delta_p)~, \label{low p}
\ee
with
\be
{\cal R}_{f}^\mu =t_f^{\mu\rho}p_{a\rho}~, \;\;{\cal R}_{i}^{\mu} =t_i^{\mu\rho}p'_{1\rho}~,
\ee
where $t_{i,f}^{\mu\rho}$ are defined in (\ref{tif}). 

Very often the soft radiation theorem is formulated  using
 the partial derivatives with respect to momenta. 
Such an expression can be derived from the original one with the help of relations 
\[
{\cal P}_{i}^{\mu}\frac{\partial}{\partial\nu}T(\nu, \Delta) =t_i^{\mu\rho}\frac{\partial}{\partial{p^\rho_a}}
T(\nu, \Delta)~, 
\]
\be
{\cal P}_{f}^{\mu}\frac{\partial}{\partial \nu} T(\nu, \Delta) =t_f^{\mu\rho}\frac{\partial}{\partial p^{'\rho}_1}
T(\nu, \Delta)~,
\label{nomentum derivatives} 
\ee 
where $\nu$ and $\Delta$ are defined in  (\ref{nu, Delta}). Let us note that the amplitude $T(\nu, \Delta)$ is defined on the mass shell of all particle momenta, therefore, generally speaking,  the partial derivatives over the components of the momenta are ill-defined because their definition requires exit from the mass shell. But due to the properties
\be
t_f^{\mu\rho} p'_{1\rho} =0, \;\; t_a^{\mu\rho} p_{a\rho} =0,
\ee
which follow from (\ref{tif}),  the amplitudes in Eq. (\ref{nomentum derivatives}) remain on the mass shell, so that such a problem does not arise.  
Using the relations (\ref{nomentum derivatives})  we obtain from (\ref{low}) 
\be
M_{\mu}=e\Bigg[j_\mu(k) +\Bigg(D^\mu_i+D^\mu_{f}\Bigg) \Bigg] T (\nu, \Delta)~, \label{low d}
\ee
where 
\be
D^\mu_f = t_f^{\mu\rho}\frac{\partial}{\partial{p^{'\rho}_{1}}}~, \;\;  
D^\mu_i = t_i^{\mu\rho}\frac{\partial}{\partial{p^\rho_{a}}}~, \label{Dif}
\ee
Using the definitions of $\nu_i = 2(p_ap_b), \nu_f =2(p'_1p'_2), \nu = \frac12(\nu_i+\nu_f), \Delta = (p_b-p'_2)^2, \Delta_p= (p_a-p'_1)^2$ one can easily check that 
\be
\Bigg(D^\mu_i+D^\mu_{f}\Bigg) T (\nu_i, \Delta) = {\cal P}_{i}^{\mu}\frac{\partial}{\partial {\nu_i}}T (\nu_i, \Delta)~, \;\; \Bigg(D^\mu_i+D^\mu_{f}\Bigg) T (\nu_f, \Delta) = {\cal P}_{f}^{\mu}\frac{\partial}{\partial {\nu_f}}T (\nu_f, \Delta)~,
\ee
\be
\Bigg(D^\mu_i+D^\mu_{f}\Bigg)T (\nu, \Delta_p) = \Bigg({\cal P}_{i}^{\mu}+{\cal P}_{f}^{\mu}\Bigg)  \frac{\partial}{\partial \nu} -2\Bigg({\cal R}_{i}^{\mu}+{\cal R}_{f}^{\mu}\Bigg)  \frac{\partial}{\partial {\Delta_p}}\Bigg] T (\nu, \Delta_p)~.
\ee
This means that all relations (\ref{low}), (\ref{low i}),(\ref{low f}),(\ref{low p}) have the same differential in momenta form (\ref{low d}) with the corresponding arguments of amplitude $T$. 

It is easy to see that the differential form (\ref{low d}) does not change if we take instead of $\nu$ and $\Delta$ any scalar variables, which are equal to them at $k=0$, which means $a\nu_i +(1-a)\nu_i$  and  $b\Delta+(1-b)\Delta_p$ with the coefficients  $0\leq a, b\leq 1$. 
 Indeed, using relation $\nu_i - \nu_f = 2\big(k((p'_a+p'_b))\big) =2\big(k((p_a+p_b))\big)$ we get
\be
j^\mu (\nu_i - \nu_f)  = \frac{p^{'\mu}_a}{(kp'_a)} 2\big(k((p'_a+p'_b))\big) -\frac{p_a^\mu}{(kp_a)} 2\big(k((p_a+p_b))\big)  = 2{\cal P}^\mu_f -2{\cal P}^\mu_i~,  \label{j nu-nu}
\ee     
while                                                                                                 
\be
D^\mu (\nu_i - \nu_f)  = \big(D^\mu_i+D^\mu_f\big) \big(2(p_ap_b) -2(p'_ap'_b)\big)= 2{\cal P}^\mu_i -2{\cal P}^\mu_f~,  \label{D nu-nu}
\ee                                                                                                     
 so that 
\be
\Big(j^\mu +D^\mu\Big)(\nu_i - \nu_f) =0~. \label{j + D  nu}
\ee     

At first sight, the result (\ref{D nu-nu}) may look confusing, since $D^\mu \sim {\cal} k^0$  and $(\nu_i - \nu_f) \sim {\cal} k^1$,  so the right-hand side  of (\ref{D nu-nu}) should also be $\sim {\cal} k^1$. 
The point is that when calculating derivatives, the vectors $p_a, p_b, p'_a,  p'_b$ are considered as being independent (not related to $k$  by momentum conservation). 
Similarly, we obtain 
\be
\big(j^\mu +D^\mu\big) (\Delta - \Delta_p )  = j^\mu 2\big(k((p'_a-p_a))\big) + \big(D^\mu_i+D^\mu_f\big)\big((p_b-p'_b)^2-(p_a-p'_a)^2\big)=  0~.  \label{j+D delta }
\ee                                                                                                                                                                                                               

Accounting for  these remarks, we get for squared modulus of the matrix element $\epsilon^\mu M_{\mu}$ with the real photon polarisation vector  $\epsilon^\mu$ 
\be
|\epsilon^\mu M_{\mu}|^2=e^2(\epsilon^\mu j_\mu(k))\Bigg[(\epsilon^\mu j_\mu(k)) +\epsilon^\mu D_\mu \Bigg] 
|T(a\nu_i +(1-a)\nu_f, b\Delta+(1-b)\Delta_p)|^2~.  \label{low sigma}
\ee
\section{The Low theorem for spin-one-half particles}
In addition to  Eq.(\ref{low}) for the scalar particles, the original paper \cite{Low:1958sn}  contains also the derivation of a similar equation  (Eq.(3.16) in  \cite{Low:1958sn}) for the case when the emitter is a spin-one-half fermion. For consideration of this case, it is convenient to use 
the same notation as in \cite{Low:1958sn}). Then, for the process    
\be
f(p_i)+\pi^0(q_i)\rightarrow f(p_f)+\pi^0(q_f) + \gamma(e, k)~,  \label{inelastic f}
\ee
where $f$ is a spin-one-half fermion  of charge $e$,   anomalous magnetic moment  $\lambda$ and mass $m$,   the result of \cite{Low:1958sn} reads
\[
M_{\mu}=\bar{u}(p_f)\Big[(e\gamma_\mu + \frac{\lambda }{2}[\gamma_\mu, \hat{k}]) \frac{1}{\hat{p}_f+\hat{k}-m} T +T \frac{1}{\hat{p}_i-\hat{k}-m}(e\gamma_\mu + \frac{\lambda }{2}[\gamma_\mu, \hat{k}])
\]
\be
+e\Big({\cal P}_{f\mu}+{\cal P}_{i\mu}\Big) \left(\frac{\partial}{\partial \nu} T\right)\Big]u(p_a) ~, \label{low fer}
\ee
where ${\cal P}_{f\mu}$  and ${\cal P}_{i\mu}$ are given by  (\ref{Pif}) and (\ref{tif})  with the replacement $p_a\rightarrow  p_i, p'_1\rightarrow  p_f$, and 
\be
T =  A(\nu, \Delta)+ \frac12(\hat{q}_i+ \hat{q}_{f})B(\nu, \Delta), \; \nu=(p_iq_i) + (p_fq_f), \;  \Delta= (q_i-q_f)^2~. \label{T fer}
\ee
We recall that the Low theorem is written via the physical momenta of the process (\ref{inelastic f}), and therefore, the criticism of \cite{Lebiedowicz:2021byo} 
repeated in \cite{Lebiedowicz:2023ell} is not justified.  

The representation (\ref{low fer}) can be rewritten  with the  ${\cal O}$ ($k$)  accuracy as 
\[
M_{\mu}=e\bar{u}(p_f)\Bigg[j_\mu T + \Bigg({\cal P}_{f\mu}+{\cal P}_{i\mu}\Bigg) \left(\frac{\partial}{\partial \nu} T\right)
\]
\be
+ \frac{[\gamma_\mu, \hat{k}])}{4(kp_f)} \Bigg(1+\frac{\lambda}{e}(\hat{p}_f+m) \Bigg)T -T \Bigg(1+\frac{\lambda}{e}(\hat{p}_i+m) \Bigg)\frac{[\gamma_\mu, \hat{k}]}{4(kp_i)}
\Bigg]u(p_i) ~, \label{low fer 1}
\ee 
where $j_\mu$ is given by (\ref{Pif})   with the replacement $p_a\rightarrow  p_i, p'_1\rightarrow  p_f$.
unpolarized, non-radiative
In Ref.\cite{Low:1958sn} the soft-photon theorem was considered on the amplitude level, and  Eq. (\ref{low fer}) is the final result of this paper. The extension of the Low theorem developed in \cite{Burnett:1967km} concerns its application to the radiative cross sections for unpolarized particles, and the main result is proof that "the first two terms of an expansion in the photon energy depend on the unpolarized, non-radiative cross-section only".

  Let us obtain the square of the matrix element summed over fermion polarizations taking the original Low theorem in its differential in momenta form, that is using
\be
j_\mu T + \big({\cal P}_{f\mu}+{\cal P}_{i\mu}\big) \left(\frac{\partial}{\partial \nu} T\right) = \Big(j_\mu  +D_\mu\big)  T~. \label{diff form fer}
\ee
Recall here that, as shown in the previous section,  in the differential form  (\ref{diff form fer}) we can set in $T$ (\ref{T fer}) instead of $\nu$ and $\Delta$ any scalar variables, which are equal to them at $k=0$. 
Using(\ref{low fer 1}) and   (\ref{diff form fer}),  we obtain for real photon polarisation vectors $\epsilon^\mu$ 
\[
\sum_{spin} |\epsilon^\mu M_{\mu}|^2=e^2\Bigg[(\epsilon j)\Big((\epsilon j)Tr\Big((\hat{p}_i +m))T(\hat{p}_f +m)\overline{T} 
\]
\[
+ Tr\Big((\hat{p}_i +m)\big((\epsilon D)T\big)(\hat{p}_f +m)\overline{T}\Big) +\Big((\hat{p}_i +m)T\big(\hat{p}_f +m)\big((\epsilon D)\overline{T}\big)\Big)
\]
\[
+ \frac{(\epsilon j)}{4(kp_f)}Tr\Bigg((\hat{p}_f +m)T(\hat{p}_i +m)\overline{T} \Big(1+\frac{\lambda}{e}(\hat{p}_f+m)\Big) [\hat{k}, \hat e ]
\]
\[
+(\hat{p}_f  +m)[\hat e, \hat{k}]\Big(1+\frac{\lambda}{e}(\hat{p}_f+m)\Big)T(\hat{p}_i +m)\overline{T}\Bigg) 
\]
\[
-\frac{(\epsilon j)}{4(kp_i)}Tr\Bigg((\hat{p}_f +m)T(\hat{p}_i +m) [\hat{k}, \hat \epsilon] \Big(1+\frac{\lambda}{e}(\hat{p}_i+m)\Big)\overline{T} 
\]
\be
+(\hat{p}_f +m)T \Big(1+\frac{\lambda}{e}(\hat{p}_i+m)\Big)[\hat \epsilon, \hat{k}](\hat{p}_i +m)\overline{T}\Bigg)~, \label{low fer 2} 
\ee
where $\sum$ means summation over the fermion polarizations. It is easy to see from (\ref{low fer 2}) that
(as already observed in \cite{Burnett:1967km})  the terms with the anomalous magnetic moment have canceled.
 Then using formulas
\[
\frac{1}{4(kp_f)}\Big([\hat{k}, \hat \epsilon](\hat p_f+m) +(\hat p_f+m)[\hat{k}, \hat \epsilon]\Big) = \epsilon_\mu t_f^{\mu\rho}\gamma_\rho = (\epsilon D)(\hat p_f  + m), 
\]
\be
\frac{1}{4(kp_i)}\Big([\hat{k}, \hat \epsilon](\hat p_i+m) +(\hat p_i+m)[\hat{k}, \hat \epsilon]\Big) = \epsilon _\mu t_i^{\mu\rho}\gamma_\rho = (\epsilon D)(\hat p_i + m), 
\ee
we get  from (\ref{low fer 2})
\be
\sum_{spin} |\epsilon^\mu M_{\mu}|^2=e^2(\epsilon  j)\Big((\epsilon j) +(\epsilon D) \Big)\sum_{spin}| M|^2~, \label{low sigma f}
\ee
where
\be
\sum_{spin}| M|^2 = 
 Tr\big((\hat{p}_i +m)T(\hat{p}_f +m)\overline{T} \big) =\sum_{spin}|\bar{u}(p_f)T u(p_i)|^2~.
  \label{Sum}
\ee
Recall that $T$ here is expressed in terms of the momenta of the radiative 
process (\ref{inelastic f}), and that we can take in $T$ (\ref{T fer}) instead of $\nu$ and $\Delta$ any scalar variables, that are
equal to them at $k=0$.\\ 
   Note that in Refs. \cite{Lebiedowicz:2023mlz, Lebiedowicz:2023ell} the soft-photon theorem for the pion-proton scattering was formulated following the same approach as in \cite{Lebiedowicz:2021byo} for the pion-pion scattering. Derivation of the result presented in \cite{Lebiedowicz:2023ell}   "involved a lengthy and complex analysis" given in \cite{Lebiedowicz:2023mlz}. The result is also 
complicated,
 so that the proof of its equivalence to the original formulation is not presented here and will be considered elsewhere.
\section{On the Burnett-Kroll extension of Low theorem}
In Ref. \cite{Low:1958sn} the soft-photon theorem was considered on the amplitude level, and  
Eqs. (\ref{low}) and (\ref{low fer}) are the final results of paper \cite{Low:1958sn} for the amplitudes of soft photon emission by spin-zero and spin-one-half particles respectively in scattering on a neutral spin-zero particle. 
Our Eqs. (\ref{low sigma}) and (\ref{low sigma f}) were obtained by a direct application of the results of \cite{Low:1958sn} to the amplitudes squared,
summed over the fermion polarizations in the second case. 
Note that our results seem to look like a confirmation (or repetition) of the conclusions of \cite{Burnett:1967km}. However, there is 
the essential difference between these results. 

For illustration, let us reproduce  the final result (Eq. (11)) of the paper \cite{Burnett:1967km}
\be
\sum_{spins}|T_\gamma(\epsilon, k, p)|^2 =\sum_a Q_a\frac{\epsilon\cdot p_a}{k\cdot p_a}\sum_b Q_b\Big[\frac{\epsilon\cdot p_b}{k\cdot p_b}-\epsilon\cdot D_b(k)\Big]\sum_{spins}|T( p')|^2, \label{BK}
\ee
where $T_\gamma(\epsilon, k, p)$ and $T( p')$ are the radiative and non-radiative amplitudes, $p_a$ and $Q_a$ are  the momentum and  charge of the  particle $a$, and $k$  and $\epsilon$  are the momentum and polarisation vector of the photon. All particles except the photon are considered as incoming, so that $\sum_a p_a =k$, $p'_a = p_a-\xi_a(k), \;\;\sum_a\xi_a(k) =k, \;\; p_a\cdot\xi_a =0$, 
\be
D_a(k) = \frac{p_a}{k\cdot p_a}k\cdot\frac{\partial}{\partial{p_a}}-\frac{\partial}{\partial{p_a}}~.
\label{D BK}
\ee

Note, that  Eqs. (\ref{low sigma}) and (\ref{low sigma f}) (as well as  Eqs. (\ref{low}) , (\ref{low fer}) ) are expressed through 
 momenta of the radiative processes, while
 in Ref. \cite{Burnett:1967km}  (see Eq.(\ref{BK}) above), the momenta of non-radiative processes are used. 
There is a large uncertainty in the choice of these momenta at the given momenta of the radiative process. An important restriction on this choice imposed in \cite{Burnett:1967km} is that "scalar invariants are the same to first order in $k$ whether expressed in terms of $p$ or $p'$ ", i.e. momenta of the radiative or non-radiative processes.  Of course, only the independent invariants are considered, because it is impossible to have simultaneously, for example,  $(p'_iq'_i) =(p_iq_i)$ and  $(p'_fq'_f) =(p_fq_f)$  since  $(p'_iq'_i) =(p'_fq'_f)$, but $(p_iq_i) =(p_fq_f) +(k((p_i+q_i))$.
It was assumed in (\ref{BK}) that $\sum_{spins}|T( p')|^2$ is expressed in invariants constructed from the momenta of the radiative process.  Note that the imposed restriction leaves a choice of momenta of the non-radiative process far from being unique.

For the case of radiation by the scalar particles, formulations of the soft-photon theorem via momenta of the radiative and non-radiative processes coincide.   
Indeed, the form (\ref{low sigma}) is in a perfect agreement with Eq. (11) of \cite{Burnett:1967km}  (see Eq.(\ref{BK}) above), because $|T(p')|^2$  expressed in invariants constructed from the momenta of the radiative process is nothing more than $|T(a\nu_i +(1-a)\nu_f, b\Delta+(1-b)\Delta_p)|^2$. The choice of different coefficients $a$ and $b$  corresponds to the choice of different scalar invariants which are the same to first order in $k$, whether expressed via $p$ or $p'$. 

The paper of Barnett and Kroll
 \cite{Burnett:1967km} was criticized in Ref.\cite{Lebiedowicz:2021byo} (see the end of Appendix B) on the grounds that "their results contain derivatives of the non-radiative amplitudes with
respect to one momentum keeping the other ones fixed". This is incorrect, as it follows from the sentence in \cite{Burnett:1967km} after Eq. (4) there: "The prescription then is that $T$ is expressed in terms of the momenta $p_a$  which are differentiated as independent variables, then evaluated with momenta $p'_a$".
 Note, that In Eq. (\ref{BK}) it is assumed that before differentiation $\sum_{spins}|T( p')|^2)$ is expressed in terms of the invariants constructed from the momenta of the radiative process.  

 We could see a weak point in
 the paper \cite{Lebiedowicz:2021byo} regarding the partial derivatives $\frac{\partial}{\partial{p_a}}$, which are generally poorly defined, as it was already noted, because their definition requires exit from the mass shell. But these derivatives enter only in the combinations  $\xi_a\cdot \frac{\partial}{\partial{p_a}}$ and $D_a(k)$, which do not require a shift from the mass shell,\\
so the only reproach is that the authors of \cite{Lebiedowicz:2021byo} did not explain this. \\
Therefore, in the case of radiation by scalar particles, the result of \cite{Lebiedowicz:2021byo}  is completely
 consistent with Eq. (\ref{low sigma}), and we have only minor comments regarding its derivation. For radiation by the spin-one-half particles, 
the situation is somewhat more complicated. In this case, the result of \cite{Lebiedowicz:2021byo}
written in the notation of Eq.(\ref{inelastic f}) takes the form 
\be
\sum |\epsilon^\mu M_{\mu}|^2=e^2(\epsilon j)\Big((\epsilon j) +(\epsilon D)\Big) \sum_{spin}| M'|^2~, 
\label{low sigma fer}
\ee

where prime labels momenta of the non-radiative process, and 
\[ 
j^\mu = \frac{p^\mu_{f}}{(p_fk)}-\frac{p_i^\mu}{(p_i k)}~,\;\;  D^\mu =t_i^{\mu\rho} \frac{\partial}{\partial{p_{i\rho}}}+t_f^{\mu\rho} \frac{\partial}{\partial{p_{f\rho}}} ~, \label{j D f}
\]		
\be
	t_i^{\mu\rho} = \frac{p_{i}^\mu}{(p_i k)} k^\rho -g^{\mu\rho}~,\;\;
	t_f^{\mu\rho} = \frac{p^{\mu}_{f}}{(p_f k)}k^\rho -g^{\mu\rho}~. \label{tif f }
\ee
\be
\sum_{spin}| M'|^2 = 
Tr\big((\hat{p}'_i +m)T(\hat{p}'_f +m)\overline{T}' \big) =\sum_{spin}|\bar{u}(p'_f)T' u(p'_i)|^2~, 
\label{Sum prime}
\ee
\be
T' = A(\nu', \Delta')+\frac12(\hat{q}'_i+\hat{q}'_2) B(\nu', \Delta')~, \nu'=2(p'_iq'_i) =2(p'_fq'_f)~, \Delta' =(q'_i-q'_f)^2 =(p'_i-p'_f)^2~. \label{T'}
\ee
The restriction that scalar invariants are the same to first order in $k$, whether expressed in terms of non-primed or primed momenta (i.e. momenta of radiative or non-radiative processes) is very important because it is needed  for the definition of
the  derivatives in (\ref{low sigma fer}). Indeed, (\ref{low sigma fer}) contains the derivatives in non-primed momenta acting on a scalar function of the primed momenta, therefore, some prescription 
for the transformation of the scalar products of the primed momenta into the scalar products of the non-primed momenta is required. \\
Let us take $\nu'_i = 2(p'_iq'_i)$ and $\Delta'_p=(p'_i-p'_f)^2$ as the independent variables, so that 
\be
\nu'_i = \nu_i, \; \; 2(p'_iq'_i) =2(p_iq_i)  ~, \;\; \Delta'_p=\Delta_p, \;\; (p'_i-p'_f)^2  = (p_i-p_f)^2~,   \label{equalities}
\ee
and take 
\be
p'_i =p_i, \;  p'_f = p_f, \;\;  q_i =q'_i+\eta_i,\;\; q_f = q'_f-\eta_f ~. 
\ee
Therefore, $T'$ becomes
\be
T' = A(\nu_i, \Delta_p) +\frac12(\hat{q}'_i+\hat{q}'_2) B(\nu_i, \Delta_p)~, 
\;\;\nu_i=2(p_iq_i)~, \;\;\Delta_p =(p_i-p_f)^2~.  \label{T' 1}
\ee

The equalities (\ref{equalities}) and the on-mass shell conditions for $q_i, q_f$  lead to the relation
\be
(\eta_i q_i) =0, \;\; (\eta_f q_f) =0, \;\; (\eta_i p_i) =0~.  \label{conditions}
\ee
Using that from momentum conservation law it follows $\eta_i+\eta_f =k$,  these conditions can be rewritten as 
\be
(\eta_i q_i) =0, \;\; (\eta_i q_f) =(kq_f), \;\; (\eta_i p_i) =0~.  \label{conditions i}
\ee
It is easy to see that this system of equations has an infinite number of solutions except for the very special choice of kinematics.  

Recall that in  Eq. (\ref{low sigma f}), in the expression for $T$ (\ref{T fer}) instead of $\nu$ and $\Delta $ we can take any variables,  which are equal to $\nu$ and $\Delta $  at $k=0$.  Let us set these  as $\nu_i$ and $\Delta_p$, i.e. substitute 
 in (\ref{Sum}) $T$ with ${\cal T}$, 
\be
{\cal T} =  A(\nu_i, \Delta_p)+ \frac12(\hat{q}_i+ \hat{q}_{f})B(\nu_i, \Delta_p)~,   \label{cal T}
\ee
so that 
\be
{\cal T} = T' +\frac12(\hat{\eta}_i- \hat{\eta}_{f})B(\nu_i, \Delta_p)~.  
\ee
Therefore, the difference between the factors in front of   $e^2(\epsilon j)^2$ in (\ref{low sigma f}) and (\ref{low sigma fer}) is (for
compactness we drop the arguments of $A$ and $B$)
\[
\sum_{spin}| M|^2 - \sum_{spin}| M'|^2 = \frac12Tr\Big((\hat{p}_f +m)(\hat{\eta}_i- \hat{\eta}_{f})B(\hat{p}_i  +m)\big(A^* + \frac12(\hat{q}'_i- \hat{q}'_{f})B^*\big) 
\]
\[
+ \frac12Tr\Big((\hat{p}_f +m) \big(A + \frac12(\hat{q}'_i- \hat{q}'_{f})B\big) (\hat{p}_i  +m)(\hat{\eta}_i- \hat{\eta}_{f})B^*\Big) = (AB^*+BA^*)2m\big((\eta_i-\eta_f)(p_i+p_f)\big)
\]
\be
 + 2BB^*\Big[\big(p_i(\eta_i-\eta_f)\big)\big(p_f(q'_i+q'_f)\big) +\big(p_f(\eta_i-\eta_f)\big)\big(p_i(q'_i+q'_f)\big)-\big((\eta_i-\eta_f)(q'_i+q'_f)\big)\big((p_ip_f) -m^2\big)
\Big]~,  \label{str}
\ee
which is not zero. Indeed, using  $\eta_f= k -\eta_i$ and the relations (\ref{conditions i}) we obtain 
\[
\sum_{spin}| M|^2 - \sum_{spin}| M'|^2  =(AB^*+BA^*)2m\big(-(k(2q_f +p_f+p_i)\big) 
\]
\be
+BB^*\big[2\nu_i\big(-(k(2q_f +p_f+p_i))\big)  -2\Delta_p\big((k(p_i+q_i))\big)\big] ~.  \label{str}
\ee
The difference  $\sum_{spin}| M|^2 - \sum_{spin}| M'|^2  $ (\ref{str}) was presumed to be zero in \cite{Burnett:1967km} because it  was written as 
\be
\frac12\Big(\big(\eta_i\frac{\partial}{\partial{q_i}} -\eta_f\frac{\partial}{\partial{q_i}}\big)Tr\big((\hat{p}_f +m)T(\hat{p}_i + +m)\overline T \big)\Big)\big|_{q_i =q'_i,q_f =q'_f }~.
\ee  
Indeed, the latter expression nullifies if we take the derivatives \underline{after} the calculation of traces with $q_i =q'_i,q_f =q'_f$ and present the result in terms of $\nu_i$ and $\Delta_p$. It was supposed in \cite{Burnett:1967km} that these operations are commutative, but this is untrue.
Such an assumption was used in \cite{Burnett:1967km}  where the
 term $\xi_b\frac{\partial}{\partial{p_b}}\sum_{spin}|T|^2$  in Eq. (10) was set  to zero. 
This circumstance on its own would be sufficient to argue that the final result of \cite{Burnett:1967km}, Eq. (11), is incorrect. But the term with derivatives in this equation is also erroneous because the same wrong assumption (commutability of taking derivatives and summation over the spins followed by the presentation of the result in terms of scalar variables expressed in momenta of the radiative process) was used in its derivation.  
Therefore, it is necessary to calculate the difference 
\be
D^\mu \sum_{spin}| M|^2 - D^\mu\sum_{spin}| M'|^2~, \label{diff D terms}
\ee
where
\be
\sum_{spin}| M|^2 = Tr\big[(\hat p_f+m)\big(A  + \frac12(\hat{q_i}+\hat{q_f})B\big)(\hat p_i+m)\big(A^*  + \frac12(\hat{q_i}+\hat{q_f})B^*\big)\big]~, \label{sum M 2}
\ee 
and $ \sum_{spin}| M'|^2 $ is obtained from (\ref{sum M 2}) by the replacement $q_i\rightarrow q'_i, q_f\rightarrow q'_f$. Note that since in the difference (\ref{diff D terms}) only ${\cal O}$ $(k^0)$ terms must be kept, and invariant amplitudes $A$ and $B$ are the same in both terms, their derivatives do not contribute to the difference so that they will be considered in what follows as constants. 

In the first term, the  derivatives are  taken \underline{before} the calculation of the trace, so that 
\[
D^\mu \sum_{spin}| M|^2 = t_f^{\mu\rho} Tr\big[\gamma_\rho\big(A  + \frac12(\hat{q'_i}+\hat{q'_f})B\big)(\hat p_i+m)\big(A^*  + \frac12(\hat{q'_i}+\hat{q'_f})B^*\big)\big]
\]
\be
+t_i^{\mu\rho} Tr \big[(\hat p_f+m)\big(A  + \frac12(\hat{q'_i}+\hat{q'_f})B\big)\gamma_\rho\big(A^*  + \frac12(\hat{q'_i}+\hat{q'_f})B^*\big)\big]~.  
\ee
Direct calculation gives
\[
D^\mu \sum_{spin}| M|^2  = 4AA^*\big(R_i^\mu +R_f^\mu\big)  +2m(AB^*+BA^*)\big(2{\cal P}_i^\mu +2{\cal P}_f^\mu -R_i^\mu -R_f^\mu\big)  
\] 
\be
+BB^*\Big[2\nu_i\big(2{\cal P}_i^\mu +2{\cal P}_f^\mu -R_i^\mu -R_f^\mu\big) +2\Delta_p\big({\cal P}_i^\mu +{\cal P}_f^\mu\big) -4\mu^2\big(R_i^\mu +R_f^\mu\big)\Big]~. \label{D}
\ee
where here $\mu$  is the mass of the scalar particle.

In the second term in  (\ref{diff D terms}) the derivatives $D^\mu$ act \underline{after} calculation of the trace and the presentation of the result in terms of $\nu_i$ and $\Delta_p =(p_i-p_f)^2$, which gives  
\[
D^\mu \sum_{spin}| M'|^2 = D^\mu\Big[2AA^*(4m^2 -\Delta_p)  +(AB^*+BA^*)2m(2\nu_i+\Delta_p) +BB^*(2\nu^2_i+2\nu_i\Delta_p +2\mu^2\Delta_p)\Big]
\]
\[
= 4AA^*\big(R_i^\mu +R_f^\mu\big) +2m(AB^*+BA^*)\big(4{\cal P}_i^\mu -2R_f^\mu -2R_i^\mu\big)  
\] 
\be
+BB^*\Big[4\nu_i\big(2{\cal P}_i^\mu  -R_i^\mu -R_f^\mu\big) +4\Delta_p{\cal P}_i^\mu  -4\mu^2\big(R_i^\mu +R_f^\mu\big)\Big]~. \label{D'}
\ee
Using (\ref{D}) and (\ref{D'}) we arrive at 
\[
D^\mu \sum_{spin}| M|^2 - D^\mu\sum_{spin}| M'|^2 =2m(AB^*+BA^*)\big(-2{\cal P}_i^\mu +2{\cal P}_f^\mu+R_f^\mu +2R_i^\mu\big)  
\] 
\be
+BB^*\Big[2\nu_i\big(2{\cal P}_f^\mu -2{\cal P}_i^\mu  +R_i^\mu +R_f^\mu\big) +2\Delta_p\big({\cal P}_f^\mu  -{\cal P}_i^\mu\big)\Big]   ~. \label{D-D'}
\ee
It is quite straightforward  to see  using (\ref{D-D'}), (\ref{str}) and explicit expressions for $j^\mu,\;\; {\cal P}_{i, f}^\mu, \;\;  R_{i, f}^\mu $  that 
\be
\Big(j^\mu +D^\mu \Big) \sum_{spin}| M|^2 -\Big(j^\mu +D^\mu \Big) \sum_{spin}| M'|^2 =0,  
\ee
which means, two errors made in \cite{Burnett:1967km} during the derivation of Eq. (\ref{BK}) as a result of the incorrect
 assumption of commutability of the two operations (taking derivatives and summation over the spins followed by the presentation of the result in terms of scalar variables written in momenta of the radiative process) cancel each other, as was shown above by the direct calculations. In reality, these calculations were not needed at all, because the cancellation can be proved using Eqs. (\ref{j + D  nu}) and (\ref{j+D delta }).  

The net outcome of our consideration is the proof of the validity of the result (\ref{low sigma fer}), 
obtained in \cite{Burnett:1967km}.  However, in our opinion, it is much more convenient to use 
the form (\ref{low sigma f}), which does not require an introduction of the artificial and ambiguously defined momenta of the non-radiative processes. 

 And finally, about the criticism in \cite{Lebiedowicz:2021byo}
concerning Ref.\cite{Lipatov:1988ii} (see Appendix A in \cite{Lebiedowicz:2021byo}).
It was based on the fact that the formula
\[
{\cal M}_\lambda\Bigg|_{Lipatov} =e\Bigg[\frac{p_{a\lambda}}{(p_a k)} -\frac{p'_{1\lambda}}{(p'_1k)}\Bigg]
{\cal M}^{(0)}(s_L, t, m^2_\pi, m^2_\pi,m^2_\pi, m^2_\pi) \label{A2}
\]
\be
-e(p_a-p_1, k)\Bigg[\frac{p_{a\lambda}}{(p_a k)} -\frac{p'_{1\lambda}}{(p'_1k)}\Bigg]\frac{\partial}{\partial t}{\cal M}^{(0)}(s_L, t, m^2_\pi, m^2_\pi,m^2_\pi, m^2_\pi)~,  \label{A2}
\ee
obtained from the equations of \cite{Lipatov:1988ii} for the case of the process (\ref{inelastic}), is  different from  Eq. (\ref{LNS}),
due to the absence of the term with  $\frac{\partial}{\partial {s_L}}{\cal M}^{(0)}$ in (\ref{A2}),  and that
the term proportional to $\frac{\partial}{\partial t}{\cal M}^{(0)}$ is different from the result of \cite{Lebiedowicz:2021byo}. 

It is easy to see that this criticism has no basis. Ref.\cite{Lipatov:1988ii} considered the high energy process in multi-Regge kinematics, assuming the Regge behaviour of the inelastic amplitude. Therefore, the  term with  $\frac{\partial}{\partial {s_L}}{\cal M}^{(0)}$  was dropped in \cite{Lipatov:1988ii} because at large energies it falls down. As for the term proportional to $\frac{\partial}{\partial t}{\cal M}^{(0)}$, it has to be different from the result of \cite{Lebiedowicz:2021byo}, since in \cite{Lipatov:1988ii} it was used
the original form (\ref{low}) of the Low theorem, which is formulated in terms of momenta of the radiative process. 
There is no term 
with $\frac{\partial}{\partial t}{\cal M}^{(0)}$ in the original formula (\ref{low}).
In (\ref{A2})
such a term appeared because in this formula
$t=\frac12 ((p_a-p'_1)^2+ (p_b-p'_2)^2)$ was used instead of  $\Delta = (p_b-p'_2)^2$ in (\ref{low}). 
It is easy to see that the second term in (\ref{A2}) is the result of the transition in (\ref{low}) from  $\Delta$ to $t$. 
\section{The Low theorem in the textbooks}
As discussed above, a conclusion in Ref.\cite{Lebiedowicz:2021byo} that the representation of soft photon emission amplitudes, given in \cite{Low:1958sn} is incomplete, is based on the assumption that it is written in terms of momenta of the
non-radiative processes that do not satisfy energy-momentum conservation. Unfortunately, such an erroneous assumption
is promoted by some popular textbooks. 
 In particular, this applies to the textbook \cite{Berestetskii:1982qgu}(and its originals in Russian \cite{BLP:1971}).
 To clarify this issue, let us first
 note  that the initial particle momenta are labeled in \cite{Berestetskii:1982qgu}
as $p_1$ and $p_2$, whereas in \cite{Lebiedowicz:2021byo} as $p_a$ and $p_b$. 
 Here for convenience, we continue using the same notation (\ref{elastic}) and (\ref{inelastic}). Then we obtain from  Eqs. (140.1), 
 (140.9) and (140.10) of \cite{Berestetskii:1982qgu} 
\be
M = (\epsilon^*)^\mu\left(M^{(-)}_\mu +M^{(0)}_\mu\right), \;
M^{(-)}_\mu = ej_\mu M^{(el)}, \;
M^{(0)}_\mu=2e\Bigg({\cal P}_{i\mu}+{\cal P}_{f\mu}\Bigg)  \frac{\partial}{\partial s} M^{(el)}~, \label{mat0}
\ee
so that 
\be
M_{\mu}=M^{(-)}_\mu +M^{(0)}_\mu=e\Bigg[j_\mu  +2\Bigg({\cal P}_{i\mu}+{\cal P}_{f\mu}\Bigg)  \frac{\partial}{\partial s}\Bigg] M^{(el)})~, \label{BLP}
\ee
where $j_\mu $ and ${\cal P}_{i,f \mu}$ are defined in (\ref{Pif}) and (\ref{tif})  and $M^{(el)}$ is defined in \cite{Berestetskii:1982qgu} as elastic scattering amplitudeite. 
Eq.(\ref{BLP}) is analogous to the original Low equation (\ref{low}), but it does not coincide with it.  In particular,(\ref{BLP}) contains an extra factor 2 in front of  $({\cal P}_{i\mu}+{\cal P}_{f\mu})$. The appearance of this factor 
 is caused by the contradictory definitions of $M^{(el)}$.  As it follows from the unnumbered equation after  Eq. (140.7) (as well as from Eq. (140.8), Eq. (140.10) and the unnumbered equation after it), $M^{(el)}$  is considered a function of $s$ and $t$. There is no problem with $t$ defined in Eq. (140.6)  as   
\be
t = (p_b-p_2')^2~, \label{t} 
\ee
since both $t$ (Eq.(\ref{t}) in \cite{Berestetskii:1982qgu}) and $\Delta$ (Eq. (\ref{nu, Delta})  in \cite{Low:1958sn}) are expressed in terms of the same momenta so that they just coincide, $\Delta\equiv t$.
It is not so neither  for the amplitudes  $M^{(el)} (s, t)$ and   $T (\nu, \Delta)$, nor for  their arguments  $s$ and $\nu$.    The point is that the amplitude $M^{(el)} (s, t)$  is not defined in \cite{Berestetskii:1982qgu}  so clearly, as  $T (\nu, \Delta)$ in  \cite{Low:1958sn}. Indeed, $M^{(el)} (s, t)$ is defined in \cite{Berestetskii:1982qgu}   by the diagram (140.5) as the amplitude of the elastic scattering process, but with the same particle momenta, as for the inelastic process (\ref{inelastic}). Moreover, there is the equality  in Eq. (140.6) in \cite{Berestetskii:1982qgu} which in the accepted notation \cite{Lebiedowicz:2021byo} reads as
\be
s=(p_a+p_b)^2 =(p_1'+p_2')^2~. \label{s}
\ee\textbf{}
This is correct for the momenta of the non-radiative process,  but not for the momenta of the process (\ref{inelastic}). Thus, the momenta of radiative and non-radiative processes are confused in \cite{Berestetskii:1982qgu}, and this leads to an error in the presentation of the theorem. 

The book \cite{Berestetskii:1982qgu} also contains a differential form of the equation (\ref{BLP}), which is claimed to be derived from  (\ref{BLP}), but this derivation is questionable.  It is based on the equation following (140.10), which reads as 
\be
2p_{b\nu} \frac{\partial}{\partial s} {\Bigg |}_t =\frac{\partial}{\partial p^\nu_a} {\Bigg |}_{p_b, p'_1, p'_2} \label{eq} 
\ee
and an analogous equation for $\frac{\partial}{\partial p'_{1'}}$. We stress again that these equations themselves are incorrect,
since the partial derivatives over components of the momenta are ill-defined, as was already discussed in Section 2 because their definition requires exit from the mass shell. It should be noted that this circumstance is not essential, since in the final formula for the differential form in \cite{Berestetskii:1982qgu}, which can be written  using expressions (\ref{Dif})  as 
\be
M_\mu=e\big(j_\mu +D_{i\mu}+D_{f\mu}\big)M^{(el)},~  \label{BLP diff}
\ee
the derivatives appear in the combinations $D_{(i, f)\mu}$ which keep momenta on the mass shell. But the disputed equations also have another, much more serious drawback. These equations are confusing because the elastic scattering momenta are related by the conservation law, and if three of them are fixed, the fourth one is also fixed, so taking a derivative with respect to it makes no sense. Actually, it is always possible to set a partial derivative with respect to a certain momentum to zero expressing the 
amplitude via other momenta. 

Due to these inaccuracies in the derivation, the equation (\ref{BLP diff}) turns out to be unequivalent to the original one (\ref{BLP}). Indeed, if we assume that $ M^{(el)}$ in Eq. (\ref{BLP diff}) is $ M^{(el)}((p_a+p_b)^2, t)$ then we obtain  (\ref{BLP}) without the term $2{\cal P}_{f\mu}$, and if we assume that $ M^{(el)}$ in Eq. (\ref{BLP diff}) is $ M^{(el)}((p'_1+p'_2)^2, t)$ then we obtain  (\ref{BLP}) without the term $2{\cal P}_{i\mu}$. Note that in the first case, we actually get the formulation
(\ref{low i}), and in the second (\ref{low f}). This means that the formulation of the Low theorem in the differential form (\ref{BLP diff}) can be considered correct if we assume that  $M^{(el)}$ in it is $M^{(el)}(s, t)$,  where   $s$ and $t$ are expressed in terms of the momenta of the radiative process. But,  in our opinion,  the formulation (\ref{BLP}) could not be justified.
\section{Conclusion}
Recently
there has been a  renewal of interest in the physics of soft photon emission and  the status  of the celebrated Low theorem \cite{Low:1958sn},
 see  for instance \cite{Lebiedowicz:2021byo},
\cite{Lebiedowicz:2023mlz}~-~\cite{Balsach:2024rkn}. 
In particular, the authors of  \cite{Lebiedowicz:2021byo} (see also \cite{Lebiedowicz:2023mlz, Lebiedowicz:2023ell})
questioned the validity of the 
Low theorem and claimed that the term of the order of $k^0$ in the expansion 
of the radiative amplitude in photon energy $k$  needs modification.
In this paper, we  demonstrate  that contrary to this claim, the Low theorem when  formulated
in terms of the final state momenta,  does not require revision. It is
shown here that the formulation  of the soft-photon theorem
  proposed in \cite{Lebiedowicz:2021byo} is in complete agreement with the 
original result of Low \cite{Low:1958sn}.
Note that though here we considered explicitly only the processes with spin zero and spin-one-half emitters,
similar arguments should hold for the soft
photon emission in other scattering processes with an arbitrary number of external charged particles
(as already mentioned in Refs.\cite {Low:1958sn, Burnett:1967km}).
 We also  
address the criticism in \cite{Lebiedowicz:2021byo} of the papers  \cite{Burnett:1967km}  and \cite{Lipatov:1988ii} and
explain why it is unsubstantiated.

It is worth emphasizing that the \underline {agreement} of the formulations does not mean their \underline{equivalence},
because different forms of presentation could differ by the values of the omitted terms.
In some sense, the situation here reminds the expansion in terms of running the coupling constant in perturbation theory.
It is well known there that when a finite number of terms in the perturbative expansion is taken,
 the definition of the coupling constant at different scales leads to different results.
  A lot of work has been devoted to this problem in QCD, where running is really important, and many recipes have been proposed for the choice of the best scale (that is, how to get the finite number of terms closer to the exact result), see for instance, the book \cite{Ioffe:2010zz}.
However, it seems to be impossible to suggest a prescription that is universally applicable to all cases.
 In the case of the soft-photon theorem, the situation is even worse, since we are talking not about the choice of one parameter, but about the choice of the amplitude in front of the expression 
which is singular in photon energy (classical current), depending on many parameters. 
Finally, in this paper, we address some inaccuracies in the discussion of the Low theorem in Ref. \cite{Burnett:1967km} and in the classic textbooks \cite{Berestetskii:1982qgu, BLP:1971}.

\end{document}